\documentclass[aip,jcp,11pt,reprint,onecolumn,floatfix,superscriptaddress,urlname]{revtex4-1}
\usepackage[utf8]{inputenc}
\usepackage{amsmath}
\usepackage{stix}
\usepackage{mhchem}
\usepackage{comment}
\usepackage{centernot}
\usepackage{graphicx}
\usepackage{booktabs}
\usepackage[dvipsnames]{xcolor}

\newcommand{\dd}{{\mathrm{d}}}
\newcommand{\pd}{\partial}
\newcommand{\pdpd}[2]{{\frac{\pd #1}{\pd #2}}}
\newcommand{\dddd}[2]{{\frac{\dd #1}{\dd #2}}}
\newcommand{\pdpddd}[3]{\pdpd{#1}{#2}\dddd{#2}{#3}}
\newcommand{\RR}{{\boldsymbol{R}}}
\newcommand{\Edmet}{{E^\mathrm{DMET}}}
\newcommand{\ddRR}[1]{{\frac{\dd #1}{\dd \RR}}}

\newcommand{\defeq}{\overset{\mathrm{def}}{=\joinrel=}}

\newcommand{\half}{{\frac{1}{2}}}
\newcommand{\ti}{\Tilde{i}}
\newcommand{\tj}{\Tilde{j}}

\newcommand{\tp}{\Tilde{p}}
\newcommand{\tq}{\Tilde{q}}
\newcommand{\tr}{\Tilde{r}}
\newcommand{\ts}{\Tilde{s}}
\newcommand{\ttt}{\Tilde{t}}
\newcommand{\ta}{\Tilde{a}}
\newcommand{\tb}{\Tilde{b}}

\newcommand{\PP}{\mathcal{P}}
\newcommand{\QQ}{\mathcal{Q}}

\newcommand{\aoeri}{{(\mu\nu|\lambda\sigma)}}
\newcommand{\emberi}{{(\tp\tq|\tr\ts)}}

\newcommand{\ECI}{E^\mathrm{CI}}

\newcommand{\CMO}{\boldsymbol{C}}
\newcommand{\bc}{\boldsymbol{c}}
\newcommand{\bM}{\boldsymbol{M}}

\newcommand {\red}[1]{#1}

\begin{document}

\title{Multi-site Reaction Dynamics Through Multi-fragment Density Matrix Embedding}

\author{Chenghan Li}
\author{Junjie Yang}
\author{Xing Zhang}
\author{Garnet Kin-Lic Chan}
\affiliation{Division of Chemistry and Chemical Engineering, California Institute of Technology, Pasadena CA 91101}

\begin{abstract}

The practical description of disordered chemical reactions, where reactions involve multiple species at multiple sites, is presently challenge using correlated electronic structure methods. Here we describe the gradient theory of multi-fragment density matrix embedding theory, which potentially provides a minimal computational framework to model such processes at the correlated electron level. We present the derivation and implementation of the gradient theory, its validation on model systems and chemical reactions \red{using density matrix embedding},  and its application to a molecular dynamics simulation of proton transport in a small water cluster, a simple example of multi-site reaction dynamics.
\end{abstract}

\maketitle

\section{Introduction}

Electron correlation plays a crucial role in the energetics of chemical reactions, but is expensive to model in large and complex systems.
Quantum embedding methods\red{\cite{sun2016quantum,lee2019projection,jones2020embedding,ma2021quantum}} present a divide and conquer approach to the electron correlation problem. They proceed by defining small embedded regions that are described accurately, surrounded by environments that are described approximately. 
Typically, the environment is either represented by some type of potential acting on the embedded system, or is modeled by explicit, but reduced, quantum degrees of freedom (i.e. a bath) that help capture the fluctuations between the embedded region and its exterior.
The density matrix embedding theory (DMET)\red{\cite{knizia2012density,knizia2013density,cui2019efficient,pham2019periodic,hermes2019multiconfigurational,tsuchimochi2015density,ye2018incremental}} is a quantum embedding method in the second category. DMET provides an exact description of embedded fragments at a mean-field level and an approximate description at the correlated level, with the capability to describe systems broken into multiple fragments of arbitrary chemical nature and size, while ensuring a continuous potential energy surface (PES).

In the current work, we derive the analytical nuclear gradient theory of DMET. Although such a gradient theory, \red{similar to the gradient theory of other embedding methods\cite{lee2019analytical,plekhanov2021calculating}}, has multiple applications, our motivation is to study reactions in settings where the ``reaction site'' is not necessarily easily defined, e.g. in a disordered process where the atoms move freely and chemical reactions can occur at multiple locations, \red{such as in nanoreactors\cite{wang2014discovering}, combustion processes\cite{zeng2020complex,ashraf2017extension}, and heterogeneous catalysis\cite{schlogl2015heterogeneous,senftle2016reaxff}}. Here, the ability of DMET to partition problems into arbitrary fragments while retaining a continuos PES, potentially positions it as a minimal theoretical framework to describe such phenomena.  Below we first describe the working equations and implementation of the DMET gradient, then illustrate its applications to some simple chemical reactions, including a reactive molecular dynamics (MD) simulation of proton transfer in a small water cluster, a prototypical example of multi-site reaction dynamics.

\section{Theory}
In the following, we adopt the Einstein summation convention, i.e., repeated indices imply summation unless otherwise noted.

\subsection{The DMET energy expression}
The theory of DMET is documented elsewhere~\cite{wouters2016practical,cui2019efficient}; we only summarize some general ideas here to establish  the energy expression and solution constraints in order to take the gradient. 

The basic idea in DMET is to decompose a large system into a set of embedded fragments or impurities (we use the terms interchangeably) coupled to baths. 
Each bath represents the
environment surrounding each fragment. As a practical approximation, DMET constructs the baths using a mean-field (i.e. Slater determinant) representation of the global ground state. 
The information in the global Slater determinant is contained in its mean-field one-particle reduced density matrix (1-RDM), $\gamma_{\mu\nu}$. Here 
$\mu, \nu, \lambda, \sigma$ index the computational atomic orbital (AO) basis; the mean-field canonical orbitals will be referred to as MOs. 
To facilitate the definition of impurities, we transform all quantities into an orthogonalized AO basis, which we term the localized orbitals (LOs), labeled $p, q, r, s$. 
As an example, the mean-field 1-RDM elements rotated into the LO basis are labelled $\gamma_{pq}$.

An impurity $I$ is defined as a subset of the full LOs. We partition the full LO basis into multiple non-overlapping impurities. For impurity $I$, the set of LOs not in $I$ defines its environment, which in DMET is later replaced by a set of bath and core orbitals. The bath of $I$ is obtained from the
singular value decomposition (SVD) of 
the  environment-impurity mean-field density matrix
\begin{align}
   \gamma_{p\centernot\in I,q\in I} = U_{p\tp} \sigma_{\tp} V_{\tp q} 
\end{align}
where $U_{p\tp}$ transforms the environment LOs into the bath orbitals. Similarly, the core orbitals for $I$ are defined from the environment-environment block of $\gamma_{pq}$, as the eigenvectors with double occupancy. Together, the set of impurity, bath, and core orbitals define the embedded impurity problem for $I$; the impurity and bath orbitals are usually referred to as the embedding orbitals (EOs) which we label with tilde (``$\sim$") labels $\tp, \tq, \tr, \ts, \ttt$; the corresponding core orbitals are labelled $\tilde{c}, \tilde{d}$. The embedding and core orbitals are, of course, different for each $I$, but where possible we suppress the additional $I$ label to reduce notation.
The transformation matrix from the LOs to the EOs of  fragment $I$ 
can be summarized as
\begin{align}
    C_{p\tp} = \begin{cases}
    \delta_{p\tp}, ~~~ p\in \text{imp}, \tp\in \text{imp} \\
    U_{p\tp}, ~~ p\in \text{env}, \tp\in \text{bath} \\
0, ~~~~\text{otherwise}
    \end{cases}
    \label{eq:Clo2eo}
\end{align}

The embedding Hamiltonian $H^I$ for impurity $I$  is defined by projecting the original Hamiltonian into the space of determinants spanned by $\{ \tp \} \otimes |\mathrm{core}\rangle$ (where $|\mathrm{core}\rangle$ is the doubly occupied core Slater determinant constructed from the core orbitals associated with $I$). Since there are different EOs for each impurity, there is a different  $H^I$ for each impurity. In second quantization, we  write 
\begin{align}
    H^I = h_{\tp\tq} a_{\tp}^\dagger a_{\tq} -\mu a_{\tp}^\dagger a_{\tp} ~(\tp\in \text{imp}) + \half \emberi a_{\tp}^\dagger a_{\tr}^\dagger a_{\ts} a_{\tq}
    \label{eq:embHam}
\end{align}
where we have included a chemical potential (see below). 
The one-electron part contains the Coulomb  and exchange contributions from the core orbitals, 
\begin{align}
\label{eq:embHam_1}
    h_{\tp \tq} &= t_{\tp \tq} + V_{\tp \tq \tilde{c} \tilde{d}}\gamma_{\tilde{c}\tilde{d}}  \nonumber \\
    &= F_{\tp\tq} - V_{\tp\tq\tr\ts} {\gamma}_{\tr\ts}
\end{align}
where $V_{\tp\tq\tr\ts}$ is the anti-symmetrized electron repulsion integral (ERI) ($(\tp\tq|\ts\tr)-\frac{1}{2}(\tp\tr|\ts\tq)$ in chemists' notation) and $t_{\tp \tq}$, $F_{\tp \tq}$ correspond to the bare one-electron and Fock operator rotated into the EO basis respectively. 
The chemical potential $\mu$ in Eq.~\ref{eq:embHam} appears only on the impurity orbitals  (i.e. not on the bath orbitals) and is the same for all fragments.  In the absence of the full self-consistency of DMET, it is needed to ensure the correct global number of electrons.

Given $H^I$, we solve for the ground-state of each embedding Hamiltonian using a quantum chemistry solver. 
The ground-state solutions of each embedded impurity problem give embedding 1-RDMs $\Gamma_{\tp\tq}^I$ and two-body reduced density matrices (2-RDMs) $\Gamma^I_{\tp\tq\tr\ts}$. 
We then evaluate the 
energy in a one-shot manner using the following recipe. We first assemble a correlated global density matrix as a weighted sum over the fragment density matrices $\Gamma^I_{\tp\tq}$:
\begin{align}
    \Gamma_{pq}' = C_{p\tp}^I \Gamma_{\tp\tq}^I w_{\tp\tq}^I C_{q\tq}^I
    \label{eq:newaodm}
\end{align}
The weight factor $w_{\tp\tq}$ is $1$ for $ \tp, \tq \in \mathrm{imp}$, $0$ for $\tp, \tq  \in  \mathrm{bath}$, and $\frac{1}{2}$ otherwise; these factors are needed to avoid double counting of energy contributions from different impurities, using the democratic partitioning convention. From the global density matrix we can solve for the chemical potential $\mu$ such that $\mathrm{tr} \ \Gamma' = N_\text{elec}$, where $N_\text{elec}$ is the desired global number of electrons. We then update the one-electron part of the embedding Hamiltonian, 
\begin{align}
    h_{\tp\tq}^{\prime} 
&   = C_{p\tp} (t_{pq} + \half V_{pqrs}\Gamma_{rs}') C_{q\tq}   - \half V_{\tp\tq\tr\ts} \Gamma_{\tr\ts}
\label{eq:hprime}
\end{align}
and compute the total DMET energy as an average over all fragments,
\begin{align}
    \Edmet = 
    h_{\tp\tq}^{\prime I} w_{\tp\tq}^I \Gamma_{\tq\tp}^I
    +\half \emberi^I w_{\tp\tq\tr\ts}^I \Gamma_{\tp\tq\tr\ts}^I 
\label{eq:Edmet}    
\end{align}
where the four-index weight $w_{\tp\tq\tr\ts}^I$ is defined as $\frac{1}{4}$ times the number of impurity labels in the set  $\{ \tp, \tq, \tr, \ts\}$. 

\subsection{The DMET energy gradient}

\red{In a variational theory, the energy gradient depends only on the AO integral derivatives.} However, from the above, we see that the DMET energy is 
not a variational expression, and thus there are additional contributions. Because the energy is not variational in the solver wavefunctions, the DMET gradient contains terms analogous to those appearing in the gradient of perturbation theories in quantum chemistry, i.e. the response of the impurity wavefunctions contributes. 
In addition,  the DMET energy depends on a set of basis transformations. The AO$\rightarrow$LO and LO$\rightarrow$EO transformations are both atomic-position-dependent, and thus their response also contributes to the energy gradient. In particular, the LO$\rightarrow$EO transformation is defined from the mean-field 1-RDM and as a result, the gradient requires the global MO response. Depending on the orthogonalization method, the AO$\rightarrow$LO transformation can be a function of AO overlap (such as for L\"{o}wdin orthogonalization\cite{lowdin1950non}) and/or the mean-field solution or additional equations (such as for intrinsic atomic orbitals\red{\cite{knizia2013intrinsic,west2013comprehensive}}). In both cases, the AO$\rightarrow$LO response depends on the AO overlap nuclear gradient.

Thus, in total, the DMET energy is a function of the set of impurity ground-state solutions $\boldsymbol{c}=\{\boldsymbol{c}_I\}$ (these could denote, e.g. the \red{configuration interaction (CI)} vectors for a full CI (FCI) solver, or the $\boldsymbol{t}$ amplitudes plus embedding MO coefficients for a coupled cluster (CC) solver), the global mean-field solution (MO coefficients) $\CMO$, and the AO integrals $\boldsymbol{A}$ (the AO 1-electron and 2-electron integrals $\boldsymbol{\mathcal{I}}$ and AO overlap integrals $\boldsymbol{S}$). The gradient (with respect to nuclear positions or any perturbation) can be expressed in terms of partial derivatives  with respect to these variables:
\begin{align}
   \frac{\dd \Edmet}{\dd \RR} = 
   \pdpddd{\Edmet}{\bc}{\RR} +
   \pdpddd{\Edmet}{\CMO}{\RR} +
   \pdpddd{\Edmet}{\boldsymbol{A}}{\RR}
   \label{eq:Efullderiv}
\end{align}
Note that $\Edmet$ does not depend on the chemical potential $\mu$ explicitly, but $\mu$ affects the energy indirectly through its effect on {$\boldsymbol{c}$}. 
The $\dd\bc/\dd\RR$ and $\dd\boldsymbol{C}/\dd\RR$ contributions are related to $\dd\boldsymbol{A}/\dd\RR$ since $\bc$ and $\boldsymbol{C}$ are implicit functions of $\boldsymbol{A}$ as defined by the solution conditions (the mean-field and impurity ground-state equations) they satisfy. Written in this form, computing the DMET gradient naively requires one to solve for $\dd\bc/\dd\RR$ and $\dd\boldsymbol{C}/\dd\RR$ for the Cartesian coordinates of each atom, i.e. $3N_\text{atom}$ times. This complexity can be greatly reduced, however, by employing the so-called $Z$-vector technique, also known as the Lagrangian method\cite{handy1984evaluation,yamaguchi1994new}.

The DMET ground-state solutions involve three sets of working equations: the solution conditions for the impurity ground-states
\begin{align}
    \boldsymbol{X} (\mu, \boldsymbol{c},\CMO,\boldsymbol{A}) = 
    \{\boldsymbol{X}^I(\mu, \boldsymbol{c}^I,\CMO,\boldsymbol{A})\} =
    \boldsymbol{0} 
\end{align}
where each $\boldsymbol{X}^I$ is the corresponding impurity solution condition (such as the CI, CC ground-state equations, etc.);
the mean-field solution condition;
\begin{align}
    {\bM}(\CMO, \boldsymbol{A}) = \boldsymbol{0}
\end{align}
and the chemical potential solution condition;
\begin{align}
    N(\bc) = \mathrm{tr} \ \Gamma' = N_\text{elec}
\end{align}
We define a Lagrangian associated with the above constraints,
\begin{align}
    \mathcal{L}  = \Edmet 
    - z (N(\bc)-N_\mathrm{elec})
    - \boldsymbol{\lambda}^\dagger \boldsymbol{X}(\mu, \bc, \CMO, \boldsymbol{A}) 
    - \boldsymbol{\alpha}^\dagger {\bM}(\CMO, \boldsymbol{A})
    \label{eq:Ldef}
\end{align}
and consider the stationary points w.r.t. all the variables.
Vanishing derivatives  of $\mathcal{L}$ w.r.t $z$, $\boldsymbol{\lambda}$ and $\boldsymbol{\alpha}$ give the working equations of DMET, while
vanishing derivatives of $\mathcal{L}$ w.r.t $\mu$, $\boldsymbol{c}$ and $\CMO$ give equations for the minimizers of the multipliers, $z_0$,  $\boldsymbol{\lambda}_0$, $\boldsymbol{\alpha}_0$ respectively,
\begin{align}
( \boldsymbol{\lambda}_0  ~ z_0) 
\begin{pmatrix}
\pdpd{\boldsymbol{X}}{\bc} & \pdpd{\boldsymbol{X}}{\mu} \\
\pdpd{N}{\bc} & 0
\end{pmatrix} = 
\Big( \pdpd{\Edmet}{\bc} ~~ 0 \Big)
\label{eq:lambdaEq}
\end{align}
and
\begin{align}
     \boldsymbol{\alpha}_0^\dagger =
    \Big(\pdpd{\Edmet}{\boldsymbol{\CMO}} - \boldsymbol{\lambda}_0^\dagger \pdpd{\boldsymbol{X}}{\boldsymbol{\CMO}}\Big)
    \Big(\pdpd{{\bM}}{\boldsymbol{\CMO}}\Big)^{-1} 
    \label{eq:alphaEq}
\end{align}
If no chemical potential is used, Eq.~\ref{eq:lambdaEq} reduces to
\begin{align}
     \boldsymbol{\lambda}_0^\dagger =
    \pdpd{\Edmet}{\boldsymbol{c}}\Big(\pdpd{\boldsymbol{X}}{\boldsymbol{c}}\Big)^{-1} \label{eq:lambdanomu}
\end{align}
which can be solved for each impurity independently due to the block-diagonal nature of $\partial{\boldsymbol{X}}/\partial\bc$, i.e., $\partial{\boldsymbol{X}}/\partial{\bc}=\text{diag}(\{\partial{\boldsymbol{X}^I}/\partial{\bc^I}\})$.

{At the stationary points of $\mathcal{L}$, $\mathcal{L} = \Edmet$ and $\dd\Edmet/\dd\RR = \dd\mathcal{L}/\dd\RR$. }
However, the nuclear gradient of $\mathcal{L}$ no longer contains the contributions from the partial derivatives w.r.t $\boldsymbol{c}$ or $\CMO$ as in Eq.~\ref{eq:Efullderiv}, as their implicit dependence on $\RR$ is accounted for by the multipliers. Thus,
\begin{align}
    \dddd{\Edmet}{\RR} =
    \pdpddd{\mathcal{L}}{\boldsymbol{A}}{\RR}
    & =  \pdpddd{\Edmet}{\boldsymbol{A}}{\RR}
    -  \boldsymbol{\lambda}_0^\dagger \pdpddd{\boldsymbol{X}}{\boldsymbol{A}}{\RR} 
    -  \boldsymbol{\alpha}_0^\dagger \pdpddd{{\bM}}{\boldsymbol{A}}{\RR} 
    \label{eq:Lderiv}
\end{align}
We note the repeated occurrence of $\Edmet-\boldsymbol{\lambda}_0^\dagger\boldsymbol{X}$ in  Eqs.~\ref{eq:alphaEq} and \ref{eq:Lderiv}, thus it is useful to define
$\mathcal{L}_\mathrm{emb} = \Edmet - \boldsymbol{\lambda}_0^{\dagger} \boldsymbol{X}$, then its derivative w.r.t. $\boldsymbol{Y} \in \{ \boldsymbol{C}, \boldsymbol{A}$\} is 
\begin{align}
   \pdpd{\mathcal{L}_\mathrm{emb}}{\boldsymbol{Y}} 
=    \pdpd{\Edmet}{\boldsymbol{Y}} - \boldsymbol{\lambda}_0^\dagger \pdpd{\boldsymbol{X}}{\boldsymbol{Y}} 
   =  \Gamma_{\tq\tp}^I w_{\tp\tq}^I\pdpd{h_{\tp\tq}^{\prime I}}{\boldsymbol{Y}} 
    +\half \Big(
    w_{\tp\tq\tr\ts}^I \Gamma_{\tp\tq\tr\ts}^I +
    D_{\tp\tq\tr\ts}^I
    \Big)
    \pdpd{\emberi^I}{\boldsymbol{Y}} 
    + D_{\tq\tp}^I \pdpd{h_{\tp\tq}^I}{\boldsymbol{Y}}
    \label{eq:dEdY}
\end{align}
which takes the form of a contraction between density matrices and Hamiltonian derivatives, 
and where we have introduced a set of auxiliary RDMs, defined as $D_{\tp\tq} \defeq -\boldsymbol{\lambda}_0^\dagger \partial \boldsymbol{X}/\partial h_{\tp\tq}$ and $D_{\tp\tq\tr\ts} \defeq -2\boldsymbol{\lambda}_0^\dagger \partial \boldsymbol{X}/\partial \emberi$, which account for all solver-specific details. In the Appendix, we give the corresponding expressions for the Hartree-Fock (HF) and FCI solvers as examples.
Choosing $\boldsymbol{Y}$ to be the AO integrals $\boldsymbol{\mathcal{I}}$ in Eq.~\ref{eq:dEdY} gives the ``skeleton" derivative of the DMET gradient, while choosing $\boldsymbol{Y}$ to be the AO overlap leads
to more complications, due to the orbital dependence on $\boldsymbol{S}$ in localization, as discussed in Appendix \ref{sec:grad_ovlp}. These two components  sum to the first two terms of Eq. \ref{eq:Lderiv}.
The remaining task is to solve for $\boldsymbol{\alpha}$ from $\partial\mathcal{L}_\mathrm{emb}/\partial\CMO$ using the coupled-perturbed Hartree-Fock (CP-HF) equations~\cite{peng1941perturbation, yamaguchi1994new}, and to consider
 the contributions from $\partial \boldsymbol{M}/\partial \boldsymbol{A}$. Further details can be found in Appendix \ref{sec:cphf}.

\section{Results and Discussion}

We first verify our derivation and implementation using some simple examples, then proceed to some prototypical reactions. In all examples, we use L\"owdin orthogonalized orbitals as our localized orbitals. We consider DMET calculations both with a chemical potential (DMET) and without a chemical potential (DMET (no $\mu$)).

\subsection{Numerical validation}

 Since DMET provides an exact embedding at the level of mean-field theory, the correctness of the DMET gradient using HF impurity solvers can be verified by comparing the DMET gradient to the gradient of a conventional HF calculation on the whole problem. The DMET gradients for correlated solvers can be validated by comparing to finite differences of the DMET energies, or by numerical integration of the analytic gradients to compare to the energies directly. 

\subsubsection{Water trimer} 

We take a water trimer  as our first example (configuration shown in Fig.~\ref{fig:3-water}A).
We used the 6-31G** basis set\cite{hariharan1973influence} and a HF impurity solver, with the DMET impurities defined as the individual atoms. The difference between the DMET energy and global HF energy is less than $10^{-11}$ a.u., and the mean absolute error (MAE) in the gradient is less than $10^{-8}$ a.u., demonstrating the correctness of the DMET energy implementation (recovering the exact mean-field embedding) as well as that of the gradient.
We next test the FCI solver using a STO-3G basis\cite{hehre1969self} on the same system and similar atomic impurities. The deviation of the analytic gradient from the finite difference gradient is shown in Fig.~\ref{fig:3-water}A.  We see that the finite difference and analytical gradients clearly agree, with the deviation between the two decreasing with the finite different displacement ($\delta x$).

\subsubsection{Hydrogen ring}

\begin{figure}
    \centering
    \includegraphics[scale=0.7]{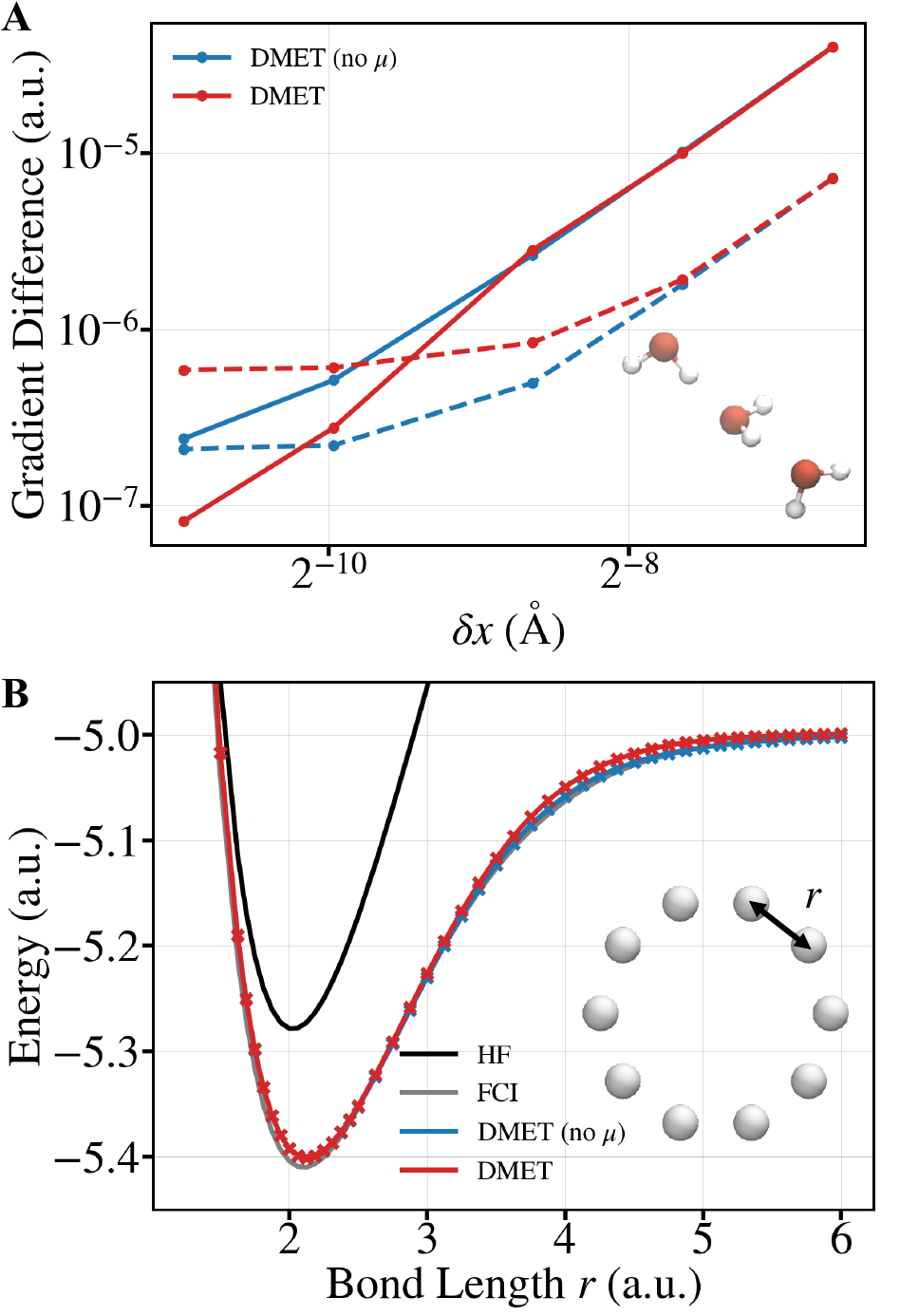}
    \caption{Numerical validation of the DMET gradient. (A) Analytic DMET nuclear gradient compared with finite difference results using different atomic position displacements ($\delta x$) in a water trimer (geometry shown). The gradient with and without chemical potential ($\mu$) fitting is shown in red and blue respectively. The solid lines indicate the \red{difference in the norm of the gradient} 
    and the dashed lines indicate the MAE. \red{The MAE tends to saturate with small $\delta x$ due to the accuracy limit of solving the response equations.} (B) PES of hydrogen ring symmetric dissociation. Solid curves represent direct calculations of the energies, while crosses represent the PES obtained by numerically integrating the DMET analytic gradient.}
    \label{fig:3-water}
\end{figure}

We next consider the symmetric dissociation of a hydrogen ring~\cite{knizia2013density}. 
We treat the $\mathrm{H}_{10}$ ring using a STO-3G basis, a FCI solver, and atomic impurities. We numerically integrate the DMET gradient along the dissociation profile to recover the PES. The correctness of the gradient is verified by the perfect match between the integrated PES and the directly calculated DMET energies (Fig.~\ref{fig:3-water}B).

Since the electron number in DMET (without $\mu$) changes only slightly along the symmetric stretching process (MAE = $\sim$0.02 with a standard deviation of $\sim$0.02) fitting $\mu$ here only slightly affects the energy.

\subsection{Diels-Alder reaction}

We next use the prototypical Diels-Alder reaction\cite{fringuelli2002diels} to illustrate using the DMET gradient  to optimize structures and locate reaction transition states.
Although the reaction in this case takes place at a single ``site'' (making 
it easy to define specific atoms and orbitals involved in the reaction) we still
treat it computationally as a multi-site problem, breaking the problem into multiple fragments in the calculation.

\begin{figure}
    \centering
    \includegraphics[scale=0.7]{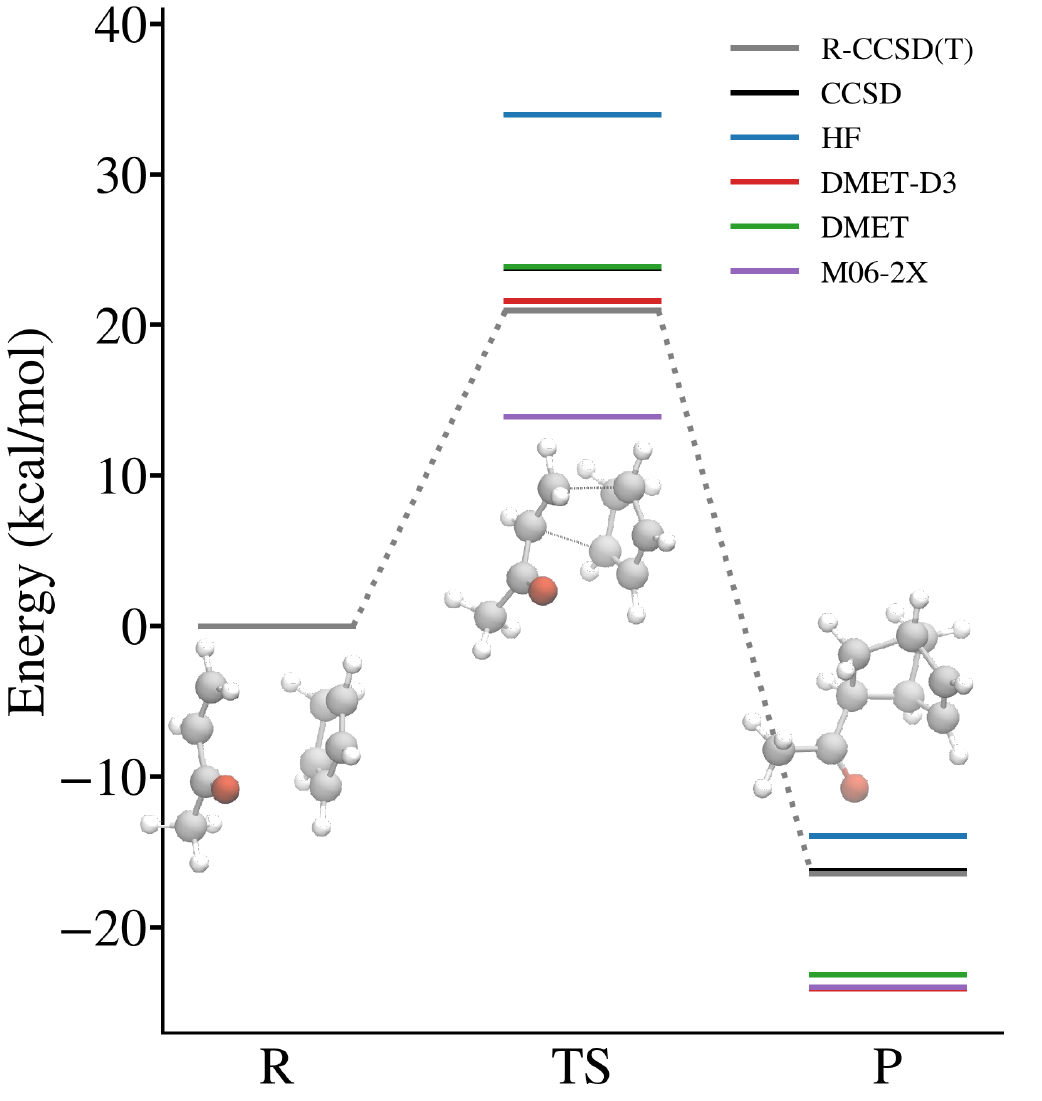}
    \caption{Optimized structures and energies of a Diels-Alder reaction. 
    R, TS, and P represent reactants, transition state and products, respectively. Energy levels for each method are aligned at R. Note that the black and green lines overlap with each other at TS. The gray and black, and red and purple overlap at P.}
    \label{fig:DA}
\end{figure}

The specific Diels-Alder reaction we choose is the addition of methyl vinyl ketone (MVK) to cyclopentadiene (CP) (see Fig.~\ref{fig:DA}), which has previously been used to test a QM/MM transition-state searching protocol\cite{yang2015qm}. We used the 6-31G basis set\cite{hehre1969self} for all atoms and each hydrogen was defined as an impurity while for heavy atoms, the atomic orbitals were split into two sets constituting two impurities: the first containing the 1s 2s 2p orbitals, and the second, the 3s 3p orbitals.
All impurity correlation problems were treated using FCI. Since DMET does not capture long-range fluctuations such as those giving rise to dispersion, we also investigated adding an empirical D3\cite{grimme2010consistent} dispersion correction (parameterized for HF) to the energy. The reactant and product states were then optimized using the DMET-D3 gradient (including the fitted chemical potential $\mu$) and the gradient was further used in a nudged elastic band (NEB)\cite{berne1998classical,henkelman2000improved,henkelman2000climbing} calculation to find the reaction transition state. The obtained transition state structure was verified via vibrational frequency analysis using the numerical Hessian computed with the DMET-D3 gradient. \red{All the optimization was performed using ASE\cite{larsen2017atomic}.}

Fig.~\ref{fig:DA} shows the DMET-D3 optimized structures and associated energies (see also Tab.~\ref{tab:DA}). In this case, DMET without fitting $\mu$ gives a large error in the total electron number (0.734 for the reactants, 0.704 at the transition state, and 0.706 for the products), thus we report only DMET results which fit the chemical potential. We use CC singles and doubles (CCSD) with renormalized triples, namely R-CCSD(T)\red{\cite{kowalski2000method}}, as the reference energy value; we note that adding the triples correction results in a relatively large shift in the reaction barrier, highlighting the open-shell character of the transition state.
After adding the empirical D3 correction, the DMET reaction barrier is similar to the R-CCSD(T) value, while DMET without D3 yields a result comparable to the CCSD barrier.  
Regardless of whether the dispersion correction is used, DMET yields a better reaction barrier compared to its underlying mean-field approach (HF), and compared to a commonly used density functional theory (DFT) approach (M06-2X\red{\cite{zhao2006new}}) (although the latter comparison is complicated by the different basis set dependence). 
In terms of the relative stability of the reactants and products, the DMET results show worse agreement with the reference R-CCSD(T) energies. This shows that using very small fragments is unfortunately not always sufficient for quantitative energetics. 

In addition to the energies, a frequency analysis at the CCSD level at the DMET-D3 transition-state shows that the DMET-D3 transition state is a good approximation to the transition state of CCSD, as \red{there is only a single large CCSD imaginary frequency (larger than 1 cm$^{-1}$) at the geometry.} The imaginary frequency obtained with the DMET-D3 numerical Hessian (562 cm$^{-1}$) is in good agreement with the imaginary frequency obtained with CCSD (581 cm$^{-1}$).

\begin{table}[]
    \centering
    \begin{tabular}{c|c|c}
    \hline\hline
    
                      & $E(\text{TS})-E(\text{R})$ & $E(\text{P})-E(\text{R})$ \\
                      \hline
        R-CCSD(T)     & 20.97 & -16.41 \\
        CCSD          & 23.78 & -16.26 \\
        HF            & 33.97 & -13.91 \\
        DMET-D3       & 21.60 & -24.05 \\
        DMET          & 23.91 & -23.12 \\
        M06-2X        & 13.93 & -23.98 \\
        \hline\hline
    \end{tabular}
    \caption{Barrier and reaction energies of a simple Diels-Alder reaction in kcal/mol. R, TS, P denote reactants, transition state, and products.}
    \label{tab:DA}
\end{table}

\subsection{Proton transport in a water cluster}

We finally consider proton transport (PT) in water as an example of a disordered chemical reaction where the reactions can involve multiple components and multiple locations, making it difficult to isolate a single reaction center to be treated by high-level quantum methods. In PT, the multi-site character arises due to the Grotthuss proton hopping mechanism\cite{de2006memoir} where an excess proton is transported by forming and breaking its surrounding  bonds with the solvent water molecules, with only minimal physical movement of the atoms. As a result, the identity of the excess proton changes dynamically and the associated reaction center cannot be tied to any specific proton or water molecule. 
\begin{figure}
    \centering
    \includegraphics[scale=0.7]{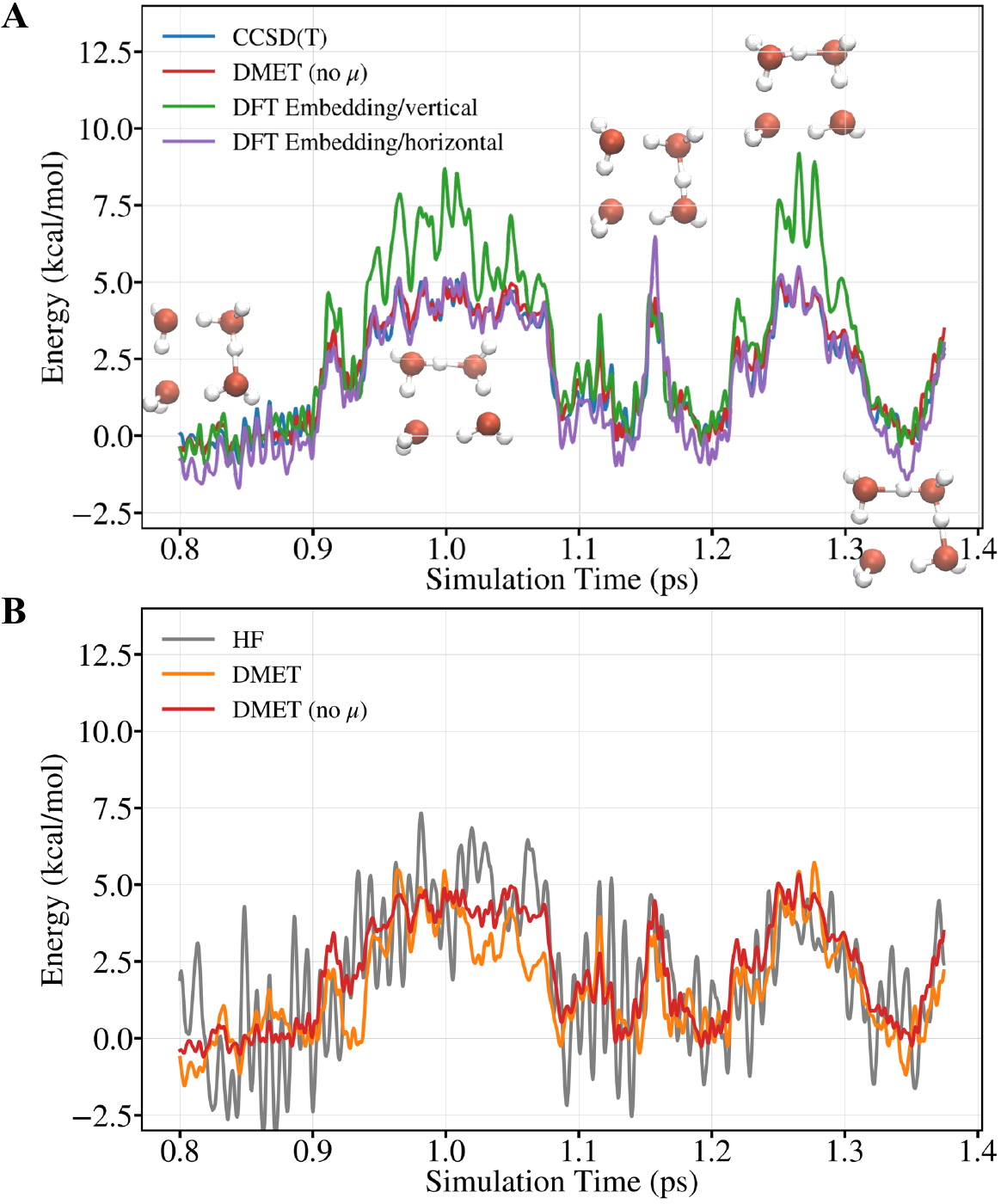}
    \caption{PES of proton transport in a water tetramer along one MD trajectory segment. 
    DFT embedding vertical means the system is partitioned into \ce{H5O2+} on the right and a water dimer on the left, while horizontal means it is partitioned into an upper \ce{H5O2+} plus a lower water dimer. The DFT embedding is performed with the projection-based wavefunction-in-DFT method using a CCSD(T) solver embedded in PBE DFT\cite{perdew1996generalized} \red{implemented in Molpro\cite{werner2020molpro}}. The DMET, HF, and DFT embedding/vertical trajectories are aligned to CCSD(T) using the \red{averaged energy in the }0.8-0.9 ps segment. The DFT embedding/horizontal trajectory is aligned using the 0.95-1.05 ps segment \red{energy average}. Lower figure shows the impact of chemical potential fitting on the DMET trajectory energy.}
    \label{fig:PT1}
\end{figure}

Fig.~\ref{fig:PT1} shows an example of such a process where one excess proton is solvated and transported in a water tetramer. The trajectory was generated by MD simulations \red{at 50K} using the DMET gradient (without fitting the chemical potential, to reduce cost) using a STO-3G basis for oxygen and the 6-31G basis for hydrogen, with each atom defined as an impurity to be treated by the FCI solver (the small oxygen basis was chosen to allow for a tractable FCI cost). The reference \red{energy curve was} generated at the CCSD(T) level \red{using MD sampled configurations}. Similar to the hydrogen ring system, the electron number fluctuated in a small range that is close to the correct number (MAE = $\sim$0.06 with standard deviation $\sim$0.002) thus chemical potential fitting is seen to change the energy only slightly. 
We used parallel-biased metadynamics\cite{pfaendtner2015efficient} to enhance the sampling of PT and the morphology of the cluster via biasing the 10 pairwise distances between the four water oxygens and a virtual atom that denotes the excess proton position\cite{li2020understanding,li2021using}.
Along these biased trajectories, the excess proton constantly changes its identity between the protons shared by the two waters on the right and those shared by the two waters on top, similar to processes seen in aqueous solution and in bio-molecules\cite{voth2006computer,berne1998classical, watkins2019proton,lee2018control,maag2021br,liu2021key,li2022proton}
making the definition of a fixed reaction center almost impossible. In such scenarios, if one wished to use a single reaction center then one might seek to  dynamically define it to follow where the reaction takes place. For example, we might change the reaction center from the right \ce{H5O2+} complex (i.e., whole system is partitioned vertically) to the upper  \ce{H5O2+} complex (i.e., partitioned horizontally) when the proton hops. 

As shown in Fig.~\ref{fig:PT1}, when two different partitioning schemes are used corresponding to the above two choices, one commonly used quantum embedding method, the projection-based wavefunction-in-DFT embedding method \cite{lee2019projection} provides an accurate description in the corresponding reaction scenarios respectively. However, relatively large errors are observed when the active excess proton goes beyond the defined fragment. Thus although each DFT embedding curve is continuous, one would have to accept a discontinuous energy jump between two different PES when changing the fragment definition to follow the excess proton identity.  
\begin{figure}
    \centering
    \includegraphics[scale=0.7]{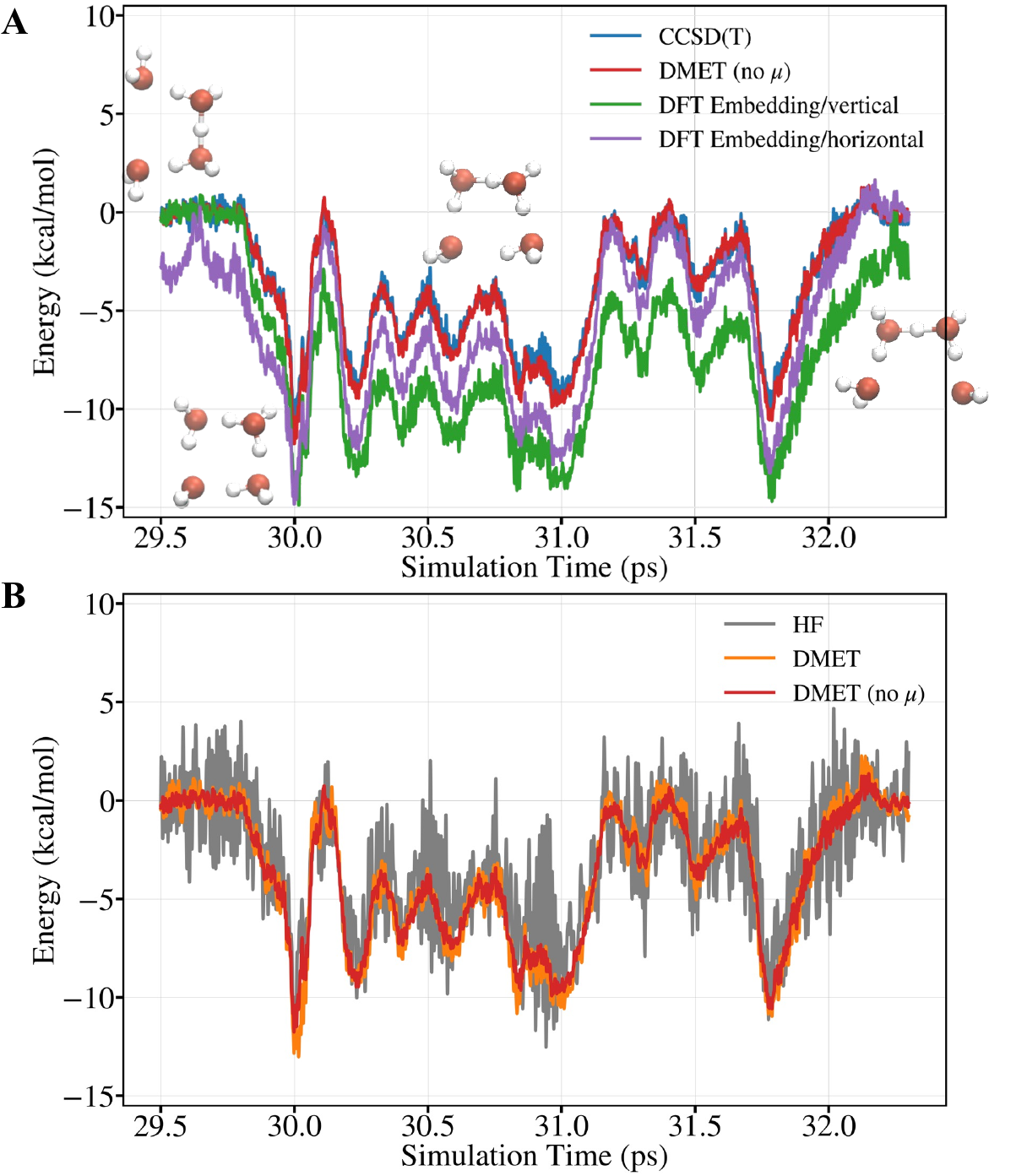}
    \caption{PES of proton transport in a water tetramer along one MD trajectory segment. Figure labels are the same as those in Fig.~\ref{fig:PT1}. The DMET, HF and DFT embedding/vertical trajectories are aligned to CCSD(T) using the \red{averaged energy in the} first 2.5 ps segment. The DFT embedding/horizontal trajectory is aligned using the last 2.5 ps segment \red{average}.}
    \label{fig:PT2}
\end{figure}
Another segment of the MD trajectory (Fig.~\ref{fig:PT2}) shows more clearly the difficulty of defining an appropriate reaction center. The excess proton remains as a shared proton between the upper two waters in the period of 30.7 ps to 32.3 ps, suggesting a horizontal partition is appropriate, while the hydrogen bond between the two lower waters is broken. We see a notable discrepancy between the DFT embedding curve and the whole-system CCSD(T) potential energy curve in the region around 30.7 ps. This is likely due to the poor description of the hydrogen bonds in the underlying mean-field (PBE) treatment in the DFT embedding. To improve this, we would then need to describe the whole problem using a high-level correlated quantum treatment, thus removing any cost savings from performing the ``quantum embedding" in the DFT embedding approach. 

In contrast, DMET produces a continuous PES that closely follows the reference CCSD(T) curve for both segments shown in Figs.~\ref{fig:PT1}, \ref{fig:PT2}, and shows a significant improvement upon the underlying mean-field (HF). This illustrates the power of the DMET bath construction which finds the most important orbitals to correlate regardless of the position of the reaction center, and demonstrates the potential of DMET as a minimal framework for multi-site reactive dynamics. An important caveat is that the quantitative accuracy of the DMET energy throughout the reaction trajectory still relies on the correlations being reasonably local. However, this appears to be the case in this simple system.

\section{Conclusion}

In this work we derived and implemented the analytic nuclear gradient of the density matrix embedding theory (DMET) and validated our  implementation via several numerical examples. In particular, we demonstrated the use of the gradient to perform geometry optimization, transition state searches and reactive molecular dynamics simulations.
In the various scenarios, we used DMET with small embedded fragments (in some cases, just a group of atomic orbitals) and observed that we could qualitatively reproduce high-level correlated calculations carried out on the full system. Importantly, the DMET potential energy surface remains continuous and thus allows for qualitatively correct reaction dynamics even for problems without a well-defined reaction center, as we demonstrated in a small water-based proton transport simulation. 

Although the DMET calculations are qualitatively reasonable across the reaction trajectory, we did not always obtain results of quantitative accuracy, due to the use of small fragments, as seen when modeling a Diels-Alder reaction. For molecular dynamics, however, small embedded fragments are desired to limit the cost of a time-step. Thus the development of empirical corrections to account for the missing correlations in DMET (perhaps along the lines of dispersion corrections) are needed. For other applications, such as single-point determinations of reaction barrier heights, more expensive impurity calculations are potentially feasible. The generalization of the gradient theory described in this work to such extensions of density matrix embedding is then an interesting future direction. 

\section{Acknowledgements}

This work, including the development of embedding gradient theory, was primarily supported by the US Department of Energy, Office of Science, via Award \red{No. DOE-SC0023318}. Additional work on the PySCF backend was supported by the US Department of Energy, Office of Science, via Award No. DOE-SC0019390. GKC is a Simons Investigator. The authors thank Dr. Zhihao Cui for helpful discussions.

\section{Appendix}
In the following derivations, we will intentionally use the component form, rather than the matrix form, of the equations where appropriate to avoid any confusion about basis and representation. The basis transformation matrices will be identified by their indexes, e.g. $C_{\mu\tp}$ is interpreted as the AO$\rightarrow$EO transformation such that $|\tp\rangle = |\mu\rangle C_{\mu\tp}$. We will also assume the orbital coefficients are real, although the equations can be straightforwardly generalized to complex-valued orbitals.

\subsection{DMET gradient terms arising from the impurity solution}
\label{sec:cpci}
The DMET energy gradient w.r.t the solution of the impurity problem is needed
 to obtain $\boldsymbol{\lambda}$ in Eq.~\ref{eq:lambdaEq}. Consider $\boldsymbol{c}^J$, the impurity solution of fragment $J$. 
Then differentiating Eq. \ref{eq:Edmet} gives
\begin{align}
    \pdpd{\Edmet}{\bc^J} 
&   =  h_{\tp\tq}^{\prime J} w_{\tp\tq}^J \pdpd{\Gamma_{\tp\tq}^J}{\bc^J} 
    + \half \emberi^J w_{\tp\tq\tr\ts}^J \pdpd{\Gamma_{\tp\tq\tr\ts}^J}{\bc^J}
    + \Gamma_{\tp\tq}^I w_{\tp\tq}^I \pdpd{h_{\tp\tq}^{\prime I}}{\bc^J} \notag \\
& \text{(no summation over $J$)} 
\label{eq:dEdc_pre}
\end{align}
where we have a sum over $I$ in the last term because $h^{\prime I}$ depends on the solutions for all fragments (see Eq.~\ref{eq:hprime}),
$\bc$, via its dependence on the updated global 1-RDM $\Gamma'$. Explicitly,
\begin{align}
\label{eq:dEdc_3}
    \Gamma_{\tp\tq}^I w_{\tp\tq}^I \pdpd{h_{\tp\tq}^{\prime I}}{\bc^J}  
&   = \Big( \half C_{p\tp}^I C_{q\tq}^I V_{pqrs} \pdpd{\Gamma_{rs}'}{\bc^J}
    - \half V_{\tp\tq\tr\ts}^J \pdpd{\Gamma_{\tr\ts}^I}{\bc^J} \delta_{IJ} \Big) \Gamma_{\tp\tq}^I w_{\tp\tq}^I  \notag \\
&   = \half \Gamma_{\tp\tq}^I w_{\tp\tq}^I C_{p\tp}^I C_{q\tq}^I V_{pqrs} 
     w_{\tr\ts}^J C_{r\tr}^J C_{s\ts}^J \pdpd{\Gamma_{\tr\ts}^J}{\bc^J}  
   - \half \Gamma_{\tp\tq}^J w_{\tp\tq}^J V_{\tp\tq\tr\ts}^J \pdpd{\Gamma_{\tr\ts}^J}{\bc^J}  \notag \\
&   = \half v_{pq}'
     w_{\tp\tq}^J C_{p\tp}^J C_{q\tq}^J \pdpd{\Gamma_{\tp\tq}^J}{\bc^J} 
   - \half \Gamma_{\tr\ts}^J w_{\tr\ts}^J V_{\tp\tq\tr\ts}^J \pdpd{\Gamma_{\tp\tq}^J}{\bc^J} \\
& \text{(no summation over $J$)}   \notag
\end{align}
where we have summed over $I$ first in $\Gamma^I_{\tp\tq}w_{\tp\tq}^IC_{p\tp}^IC_{q\tq}^I$, used Eq.~\ref{eq:newaodm}, and performed some dummy variable changes, and
defined the effective potential $v_{rs}'=\Gamma_{pq}'V_{pqrs}$.
 Inserting back into Eq.~\ref{eq:dEdc_pre}, 
\begin{align}
\label{eq:dEdc_general}
    \pdpd{\Edmet}{\bc^J}
&   = \Big [ 
    h_{\tp\tq}^{\prime J} w_{\tp\tq}^J
   - \half \Gamma_{\tr\ts}^J w_{\tr\ts}^J V_{\tp\tq\tr\ts}^J 
   + \half v_{pq}' w_{\tp\tq}^J C_{p\tp}^J C_{q\tq}^J 
    \Big]  \pdpd{\Gamma_{\tp\tq}^J}{\bc^J} 
   + \half \emberi^J w_{\tp\tq\tr\ts}^J \pdpd{\Gamma_{\tp\tq\tr\ts}^J}{\bc^J}  \notag \\
&   \defeq h^{\prime\prime J}_{\tp\tq}  \pdpd{\Gamma_{\tp\tq}^J}{\bc^J} + \half \emberi^J w_{\tp\tq\tr\ts}^J \pdpd{\Gamma_{\tp\tq\tr\ts}^J}{\bc^J} \\
& \text{(no summation over $J$)}  \notag
\end{align}
where the terms in the square brackets define $h_{ij}''^J$. 

In the case of no chemical potential fitting, $\boldsymbol{\lambda}$ is directly obtained after a contraction with $(\partial{\bM}/\partial{\bc})^{-1}$ (where the contraction is performed by solving the response equation of the underlying solver method, such as coupled perturbed CI (CP-CI)\cite{yamaguchi1994new} for a CI solver, or the $\Lambda$-equations for CC solvers\cite{shavitt2009many}).
If chemical potential fitting is performed, the solver responses of different impurities are connected via a common chemical potential. Thus their $\boldsymbol{\lambda}$'s are solved for at the same time as $z$, the multiplier associated with the electron number constraint (Eq.~\ref{eq:lambdaEq}).

\subsubsection{Hartree-Fock solver response}
\label{sec:hf_solver}
The Hartree-Fock impurity solver response enters into the DMET gradient theory in two ways: (1) it is the starting point for the response theory of more sophisticated post-Hartree-Fock solvers such as CC, truncated CI and complete active space (CAS)-CI, (2) it enters directly if Hartree-Fock is used to treat some of the impurities while others are treated by correlated methods. 
The DMET energy gradient w.r.t the HF solution of the impurity problem (Eq.~\ref{eq:dEdc_general}) is (we consider $(\partial\Edmet/\partial\boldsymbol{C})\boldsymbol{C}$ instead of $\partial\Edmet/\partial\boldsymbol{C}$ to connect to the well-known CP-HF equation)
\begin{align}
    \pdpd{\Edmet}{C_{\tp\ti}} C_{\tp\ta}
    = \pdpd{\Edmet}{\Gamma_{\tp\tq}} \pdpd{\Gamma_{\tp\tq}}{C_{\tr\ti}} C_{\tr\ta}
    = (f'_{\ta\ti}+f'_{\ti\ta})
    \label{eq:dEdc_HF}
\end{align}
In the above, we use $\ti$ and $\ta$ to label the Hartree-Fock impurity orbitals, i.e. the MOs in the embedding space ($\ti$, $\tj$ will be used for occupied and $\ta$, $\tb$ for virtual orbitals) while $\tp$, $\tq$, $\tr$ and $\ts$, consistent with the main text, denote embedding orbitals, i.e., the computational basis in the embedding space. The embedding MO coefficient matrix is thus written as $C_{\tp\ti}$ or $C_{\tp\ta}$, and the embedding ``Fock" matrix is defined as
\begin{align}
    f'_{\tp\tq}
    = \pdpd{\Edmet}{\Gamma_{\tp\tq}}
    = h''_{\tp\tq} + V_{\tp\tq\tr\ts} w_{\tp\tq\tr\ts}\Gamma_{\tr\ts}
\end{align}

The CP-HF equation is usually written as 
\begin{align}
    \mathcal{A}_{{\tb}{\tj},{\ta}{\ti}} U_{{\ta}{\ti}}^\RR = B^\RR_{{\tb}{\tj}}
    \label{eq:cphf}
\end{align}
where $\mathcal{A}$ and $B^\RR$ in our language are $\mathcal{A}=-\pdpd{\boldsymbol{X}}{\CMO} \CMO$ and $B^\RR=\pdpddd{\boldsymbol{X}}{\boldsymbol{A}}{\RR}$ with $X_{\ta\ti}=f_{\ta\ti}=0$ the Hartree-Fock equation of the embedding problem. The solution $\boldsymbol{U}^\RR$ is related to MO response via  $\CMO^{-1}(\dd{\CMO}/\dd{\RR})$.
Comparing to Eq.~\ref{eq:lambdanomu}, if no $\mu$-fit is performed, $\boldsymbol{\lambda}$ for the HF solver is obtained from the CP-HF equation by replacing $B^\RR$ with $f'$, i.e.,
\begin{align}
    -\lambda_{\ta\ti} = 2 f'_{\tb\tj} \mathcal{A}^{-1}_{\tb\tj,\ta\ti}
\end{align}
The auxiliary RDMs for the HF solver (defined after Eq.~\ref{eq:dEdY}) are then obtained by contracting with $\pdpd{\boldsymbol{X}}{h_{\tp\tq}}$ and $\pdpd{\boldsymbol{X}}{\emberi}$ 
\begin{align}
    D_{\tp\tq} &= \lambda_{\ta\ti} C_{\tp\ta} C_{\tq\ti} \\
    D_{\tp\tq\tr\ts} &= 
    D_{\tp\tq} \Gamma_{\tr\ts} - \half D_{\tp\ts}\Gamma_{\tq\tr} +
    \Gamma_{\tp\tq} D_{\tr\ts} - \half \Gamma_{\tp\ts}D_{\tq\tr}
\end{align}
If a $\mu$-\red{fitting} is performed, the above auxiliary RDM expressions still hold but $\boldsymbol{\lambda}$ should be obtained from Eq.~\ref{eq:lambdaEq} using all the impurity response equations simultaneously. The other information needed in Eq.~\ref{eq:lambdaEq} is
\begin{align}
    \pdpd{N}{C_{\tp\ti}} = 
    \begin{cases}
    2 C_{\tp\ti} , ~~~~\tp \in \text{imp} \\
    0,~~~~~~~~\text{otherwise}
    \end{cases}
\end{align}
and 
\begin{align}
    \pdpd{X_{\ta\ti}}{\mu} = - \sum_{\tp\in\text{imp}} C_{\tp\ta} C_{\tp\ti} 
\end{align}

\subsubsection{FCI solver response}
The FCI vector $c_\PP$ is the solution of the equations,
\begin{align}
    H_{\PP\QQ} c_\QQ & = \ECI c_\PP \notag \\
    \sum_\PP c_\PP^2 & = 1
    \label{eq:fciEq}
\end{align}
where we use $\PP$ and $\QQ$ to label an electronic configuration (e.g. a determinant).
The RDMs are constructed from $c_\PP$ via
\begin{align}
    \Gamma_{\tq\tp} =  c_\PP c_\QQ \Gamma^\mathcal{PQ}_{\tq\tp}  \\
    \Gamma_{\tp\tq\tr\ts} = c_\PP c_\QQ \Gamma^\mathcal{PQ}_{\tp\tq\tr\ts}
\end{align}
where $\Gamma^\mathcal{PQ}_{\tq\tp}=\langle\Phi_\mathcal{P}|a^\dagger_{\tp} a_{\tq}|\Phi_\mathcal{Q}\rangle$ and $\Gamma^\mathcal{PQ}_{\tp\tq\tr\ts}=\langle\Phi_\mathcal{P}|a^\dagger_{\tp}a^\dagger_{\tr}a_{\ts}a_{\tq}|\Phi_\mathcal{Q}\rangle$. 
Note that since the FCI solution only depends on the space of $\PP$, we can perform the CI calculation directly in the EO basis without the need to perform an embedded HF calculation (to determine MOs). Then Eq.~\ref{eq:dEdc_general} becomes
\begin{align}
 \pdpd{\Edmet}{c_\PP} 
&   =  2c_\QQ  \Big(h''_{\tp\tq} \Gamma_{\tp\tq}^{\QQ\PP} + \half \emberi w_{\tp\tq\tr\ts}\Gamma_{\tp\tq\tr\ts}^{\QQ\PP} \Big)  \defeq b_\PP
\label{eq:dEdc_fci}
\end{align}
which has the form of a CI Hamiltonian contracted with the CI vector $2c_\QQ$ to give a new CI vector. 
Note that if there are no weights $w$ and if $h$ instead of $h'$ is used to compute the energy, Eq.~\ref{eq:dEdc_fci} simply becomes $2 c_\QQ H_{\PP\QQ}=2 \ECI c_\PP$, which vanishes when dotted with $\mathrm{d}c_\PP/\mathrm{d}\RR$ due to the orthogonality between $\bc$ and $\mathrm{d}\bc/\mathrm{d}\RR$. 
It is obvious that such a Hellmann-Feynman condition is not fulfilled, and this is why the CI response is needed for the DMET gradient even though CI is a variational method. Differentiating the CI equation (Eq.~\ref{eq:fciEq}) w.r.t $c_\PP$ and after some linear algebra, we have
\begin{align}
    \pdpd{X_\QQ}{c_\PP} 
    = H_{\QQ\PP} - \ECI\delta_{\PP\QQ} + 2 c_{\PP} c_{\QQ}
    \defeq \mathcal{X}_{\PP\QQ}
\end{align}
In the case of no $\mu$-fitting, we obtain 
\begin{align}
    \lambda_\QQ = b_\PP  \mathcal{X}^{-1}_{\PP\QQ}
    \label{eq:fci_lambda}
\end{align}
i.e., $\boldsymbol{\lambda}$ is the solution of a CP-CI equation with $b_\PP$. 
Given that
\begin{align}
    \pdpd{X_{\PP}}{h_{\tp\tq}} &= \Gamma_{\tq\tp}^{\PP\QQ}c_\QQ - \Gamma_{\tq\tp} c_\PP \\
    \pdpd{X_{\PP}}{\emberi} &= \half (\Gamma_{\tp\tq\tr\ts}^{\PP\QQ}c_\QQ - \Gamma_{\tp\tq\tr\ts} c_\PP)
\end{align}
the auxiliary RDMs defined in Eq.~\ref{eq:dEdY} become
\begin{align}
D_{\tq\tp} = \lambda_\PP c_\PP \gamma_{\tq\tp} - \lambda_\PP c_\QQ \gamma_{\tq\tp}^{\PP\QQ} \\
D_{\tp\tq\tr\ts} = \lambda_\PP c_\PP \Gamma_{\tp\tq\tr\ts} - \lambda_\PP c_\QQ \Gamma_{\tp\tq\tr\ts}^{\PP\QQ}
\end{align}
To solve for $\boldsymbol{\lambda}$ with $\mu$-fitting from Eq.~\ref{eq:lambdaEq}, we also need
\begin{align}
    \pdpd{N}{c_\PP} = \sum_{\tp\in\text{imp}} 2 c_\PP \delta_{\tp\in\PP}
\end{align}
where $\delta_{\tp\in\PP}=1$ if and only if $\tp$ is occupied in configuration $\PP$, and
\begin{align}
    \pdpd{X_\PP}{\mu} = 
    \sum_{\tp\in\text{imp}} \Gamma_{\tp\tp}c_\PP - \Gamma_{\tp\tp}^{\PP\QQ} c_\QQ
\end{align}

\subsection{Orbital transformation gradient}
\label{sec:orbital_response}
The DMET gradient includes contributions from the response of the orbital transformations. A position change induces both AO$\rightarrow$LO and LO$\rightarrow$EO changes that contribute to a change in the energy. This response is encoded in the derivative of $\mathcal{L}_\text{emb}$ w.r.t the AO$\rightarrow$LO and LO$\rightarrow$EO coefficients, while both of them, via the chain rule, are related to the derivative  $ \pd{\mathcal{L}_\mathrm{emb}}/\pd{C_{\mu\tp}}$. Following our conventions for basis indexing, 
$C_{\mu\tp}= C_{\mu p} C_{p\tp}$ is the transformation matrix from AO to EO, and $C_{\mu p}$ represents the AO$\rightarrow$LO and $C_{p\tp}$ represents the LO$\rightarrow$EO transformation as defined in Eq.~\ref{eq:Clo2eo}.

We write the three components of $\pd{\mathcal{L}_\mathrm{emb}}/\pd{C_{\mu\tp}}$ appearing in
Eq.~\ref{eq:dEdY} as
\begin{align}
\label{eq:dEdgamma_1}
    D_{\tq\tp} \pdpd{h_{\tp\tq}}{C_{\mu\ttt}}
&   = 2 D_{\tq\ttt} F_{\mu q} C_{q\tq} 
    - 2D_{\tq\tp} V_{\tp\tq\ttt\ts} C_{q\ts}  \gamma_{\nu q}S_{\mu\nu} \notag \\
&   -\Big( D_{\tq\ttt} V_{\mu\tq\tr\ts} \gamma_{\tr\ts}
  + D_{\tq\ttt} V_{\tq\mu\tr\ts} \gamma_{\tr\ts}
  + D_{\tq\tr} V_{\tr\tq\mu\ts} \gamma_{\ttt\ts}
  + D_{\tq\ts} V_{\ts\tq\tr\mu} \gamma_{\tr\ttt} \Big) \notag \\
&  = 2\Big(
    D_{\tq\ttt} F_{\mu\tq}  
    - D_{\tq\tp} V_{\tp\tq\ttt\ts} \gamma_{\nu\ts} S_{\mu\nu}
    - D_{\tq\ttt} V_{\mu\tq\tr\ts} \gamma_{\tr\ts}
    - \gamma_{\tq\ttt} V_{\mu\tq\tr\ts} D_{\tr\ts}
    \Big) 
\end{align}
\begin{align}
    \half (D_{\tp\tq\tr\ts}+\Gamma_{\tp\tq\tr\ts} w_{\tp\tq\tr\ts}) \pdpd{\emberi}{C_{\mu\ttt}}
&   = 2 (D_{\ttt\tq\tr\ts}+\Gamma_{\ttt\tq\tr\ts} w_{\ttt\tq\tr\ts}) (\mu\tq|\tr\ts) 
      \label{eq:dEdgamma_2}
\end{align}
\begin{align}
\label{eq:dEdgamma_3}
    \Gamma_{\tq\tp}w_{\tp\tq}\pdpd{h'_{\tp\tq}}{C_{\mu\ttt}}
&   = 2\Gamma_{\tq\ttt}w_{\ttt\tq} (t_{\mu q} +v'_{\mu q}) C_{q\tq} \notag \\
&   - \half 
    \Big(\Gamma_{\tq\ttt}w_{\ttt\tq} V_{\mu\tq\tr\ts} \Gamma_{\tr\ts} 
  + \Gamma_{\tq\ttt}w_{\ttt\tq} V_{\tq \mu\tr\ts} \Gamma_{\tr\ts}
  + \Gamma_{\tq\tr}w_{\tq\tr} V_{\tr\tq \mu\ts} \Gamma_{\ttt\ts}
  + \Gamma_{\tq\ts}w_{\tq\ts} V_{\ts\tq\tr \mu} \Gamma_{\tr\ttt} \Big) \notag \\
&   = 2\Gamma_{\tq\ttt}w_{\ttt\tq} (t_{\mu q} +V_{\mu q rs}\Gamma'_{rs}) C_{q\tq} 
- \Gamma_{\tq\ttt} w_{\tq\ttt} V_{\mu\tq\tr\ts} \Gamma_{\tr\ts}
- \Gamma_{\tq\ttt} V_{\mu\tq\tr\ts} \Gamma_{\tr\ts}w_{\tr\ts}
\end{align}

\subsection{DMET gradient contribution from the global mean-field}
\label{sec:cphf}

The DMET energy response w.r.t the mean-field solution ($\pd\mathcal{L}_\mathrm{emb}/\pd\CMO$) is required to solve for $\boldsymbol{\alpha}$ (Eq.~\ref{eq:alphaEq}). We label the global occupied MOs by $i$, $j$ and the virtuals by $a$, $b$. $\CMO$ satisfies
\begin{align}
    C_{\mu i} F_{\mu\nu} C_{\nu a} = 0
\end{align}

Similar to the treatment of the HF solver in section \ref{sec:hf_solver}, it is more convenient to consider $\pd\mathcal{L}_\mathrm{emb}/\pd\gamma_{\mu\nu}\defeq \mathcal{F}_{\mu\nu}$ than $\pd\mathcal{L}_\mathrm{emb}/\pd\CMO$.
Assuming we use L\"{o}wdin orthogonalization to obtain the LOs, then there is no mean-field contribution to the AO$\rightarrow$LO transformation gradient. As such, $h'_{\tp\tq}$ and $\emberi$ depend on $\gamma_{\mu\nu}$ only through the LO$\rightarrow$EO transformation $C_{p\tp}$, which is obtained from the SVD of $\gamma_{pq}$. On the other hand, $h_{\tp\tq}$ not only depends on $C_{p\tp}$ due to the basis transformation but also has the extra ``skeleton" dependency on $\gamma$, due to its explicit appearance in the global Fock matrix used to define $h_{\tp\tq}$ (Eq.~\ref{eq:embHam_1}).

We first consider the ``non-skeleton" part of $\mathcal{F}_{\mu\nu}$, arising from the LO$\rightarrow$EO transformation, expressed as a chain of partial derivatives,
\begin{align}
\label{eq:Fkl}
    \pdpd{\mathcal{L}_\mathrm{emb}}{C_{\mu\tp}^I}
    \pdpd{C_{\mu\tp}^I}{\gamma_{rs}} \pdpd{\gamma_{rs}}{\gamma_{\lambda\sigma}}
    = \pdpd{\mathcal{L}_\mathrm{emb}}{C_{\mu\tp}^I}
    C_{\mu p} \pdpd{C_{p\tp}^I}{\gamma_{rs}} C_{r\lambda} C_{s\sigma} \notag \\
     \text{ ($r\in I$, $s, p\centernot\in I$ and $\tp\in$ bath)}
\end{align}
The $r$ and $s$ indices in Eq.~\ref{eq:Fkl} are restricted because the bath is constructed from the env-imp block of $\gamma$. The $p$ and $\tp$ indices are restricted to be in the environment and bath since the other elements of the LO$\rightarrow$EO transformation matrix are $0$ or $1$ (Eq.~\ref{eq:Clo2eo}) and do not contribute to the derivative. 
The partial derivative $\partial C^I_{p\tp}/\partial \gamma_{rs}$ corresponds to the standard expression for the derivative of the SVD vectors w.r.t the SVD input matrix, and the prefactor $\partial\mathcal{L}_\text{emb}/\partial C_{\mu\tp}$ in Eq.~\ref{eq:Fkl} is given in Appendix \ref{sec:orbital_response}.

As discussed earlier, the ``skeleton" part of $\mathcal{F}_{\mu\nu}$ has the contribution from $h$ only:
\begin{align}
    D_{\tq\tp}^I h_{\tp\tq}^{I [\gamma_{\mu\nu}]}
    = D_{\tq\tp}^I  V_{\tp\tq\mu\nu}^I - D_{\tq\tp}^I V_{\tp\tq\tr\ts}^I C_{\tr\mu}^I C_{\ts\nu}^I
    \label{eq:dEdgamma_skel}
\end{align}
Adding Eq.~\ref{eq:Fkl} and Eq.~\ref{eq:dEdgamma_skel} completes the DMET ``Fock" matrix $\mathcal{F}_{\mu\nu}$, and when combined with the CP-HF equation, we obtain
the third term of the DMET gradient (Eq.~\ref{eq:Lderiv}) as
\begin{align}
    -\boldsymbol{\alpha}_0^\dagger \pdpddd{\bM}{\boldsymbol{A}}{\RR}
&   = 2\mathcal{F}_{ai}\mathcal{A}^{-1}_{ai,bj} B^\RR_{bj}
    - \mathcal{F}_{ij}S_{ij}^{[\RR]}
    \label{eq:Lderiv_3}
\end{align}
where $S^{[\RR]}$ is the ``skeleton" nuclear gradient of the MO overlap matrix, i.e., $S_{\mu\nu}^\RR C_{\mu i}C_{\nu j}$, and the expressions for $\mathcal{A}$ and $B^\RR$ can be found in the CP-HF literature, such as Ref.~\cite{yamaguchi1994new}.

\subsection{DMET gradient arising from the AO overlap and AO integrals}
\label{sec:grad_ovlp}
We consider the first two terms of the DMET gradient (Eq.~\ref{eq:Lderiv}) to be composed of two contributions:
\begin{align}
\label{eq:dLdA_decomp}
   \pdpd{\mathcal{L}_\text{emb}}{\boldsymbol{A}} \boldsymbol{A}^\RR 
=  \pdpd{\mathcal{L}_\text{emb}}{\boldsymbol{\mathcal{I}}} \boldsymbol{\mathcal{I}}^\RR
+  \pdpd{\mathcal{L}_\text{emb}}{S_{\mu\nu}} S_{\mu\nu}^\RR
\end{align}
The first term comes from computing the embedding Lagrangian using AO integral derivatives ($t_{\mu\nu}^\RR$ and $\aoeri^\RR$) in place of the original integrals:
\begin{align}
    \pdpd{\mathcal{L}_\mathrm{emb}}{\boldsymbol{\mathcal{I}}} \boldsymbol{\mathcal{I}}^\RR
   =  \Gamma_{\tq\tp}^I w_{\tp\tq}^I h_{\tp\tq}^{\prime I[\RR]}
    +\half \Big(
    w_{\tp\tq\tr\ts}^I \Gamma_{\tp\tq\tr\ts}^I +
    D_{\tp\tq\tr\ts}^I
    \Big)
    \big[\emberi^{I}\big]^{[\RR]}
    + D_{\tq\tp}^I h_{\tp\tq}^{I[\RR]}
    \label{eq:dLdI}
\end{align}
Regarding the second term in Eq.~\ref{eq:dLdA_decomp}, the Lagrangian $\mathcal{L}_\text{emb}$ depends on $S_{\mu\nu}$ in two ways: 
(1) from the explicit appearance of $S_{\mu\nu}$ in $h_{\tp\tq}$ when transforming the AO 1-RDM into the EO basis using $C_{\tp\mu}=S_{\mu\nu}C_{\nu\tp}$, 
(2) and the implicit dependency on $S_{\mu\nu}$ of the LOs.
The first contribution, which can be regarded as a ``skeleton" derivative, is simply
\begin{align}
D_{\tq\tp}^I h_{\tp\tq}^{I[S_{\mu\nu}]} S_{\mu\nu}^\RR
    = -2 D_{\tq\tp}^I V_{\tp\tq\tr\ts}^I\gamma_{\mu\lambda} C_{\ts\lambda}^I C_{\nu\tr}^I S_{\mu\nu}^\RR
    \label{eq:dLdS_1}
\end{align}
The ``non-skeleton" contribution can be decomposed further into two components related to $C_{\mu p}$ and $C_{p\tp}$ respectively:
\begin{align}
\label{eq:dLdS_decomp}
    \pdpd{\mathcal{L}_\text{emb}}{C_{p\tp}^I}
    \pdpd{C_{p\tp}^I}{S_{\mu\nu}} S_{\mu\nu}^\RR
+   \pdpd{\mathcal{L}_\text{emb}}{C_{\mu p}}
    \pdpd{C_{\mu p}}{S_{\lambda\sigma}} S_{\lambda\sigma}^\RR
\end{align}
of which the $C_{p\tp}$ part of gradient can be computed from the chain rule,
\begin{align}
 \pdpd{\mathcal{L}_\mathrm{emb}}{C_{p\tp}^I}
 \pdpd{C_{p\tp}^I}{\gamma_{rs}} \pdpd{\gamma_{rs}}{S_{\mu\nu}} S_{\mu\nu}^\RR
 = \pdpd{\mathcal{L}_\text{emb}}{C_{\lambda \tp}^I} C_{\lambda p} \pdpd{C_{\lambda\tp}^I}{\gamma_{rs}}\Big(\ddRR{C_{r\mu}}C_{s\nu} + C_{s\mu}\ddRR{C_{r\nu}} \Big) \gamma_{\mu\nu}
 \label{eq:dLdS_2}
\end{align}
where the prefactor before the parentheses is the same as the prefactor in Eq.~\ref{eq:Fkl} and the same orbital restrictions are also applied here. The $C_{p\mu}$ gradient is related to the AO$\rightarrow$LO gradient via
\begin{align}
    \ddRR{C_{p\nu}}=\ddRR{C_{\mu p}} S_{\nu\mu} + C_{\mu p} S_{\nu\mu}^\RR  = \pdpd{C_{\mu p}}{S_{\lambda\sigma}} S_{\lambda\sigma}^\RR S_{\nu\mu}  + C_{\mu p} S_{\nu\mu}^\RR
\end{align}
Since we assume we are using L\"{o}wdin orthogonalization, $C_{\mu p}$ is a pure function of the $S_{\mu\nu}$ eigenvalues and eigenvectors, and thus $\partial C_{\mu p}/\partial S_{\lambda\sigma}$ 
can be rewritten using well-known expressions for matrix eigenvalue and eigenvector derivatives.
Again using the chain rule, the $C_{\mu p}$-related overlap response (second term in Eq.~\ref{eq:dLdS_decomp}) can be written as,
\begin{align}
   \pdpd{\mathcal{L}_\mathrm{emb}}{C_{\mu p}}
  \pdpd{C_{\mu p}}{S_{\lambda\sigma}} S_{\lambda\sigma}^\RR
  = \pdpd{\mathcal{L}_\mathrm{emb}}{C_{\mu \tp}^I} C_{p\tp}^I \pdpd{C_{\mu p}}{S_{\lambda\sigma}} S_{\lambda\sigma}^\RR
  \label{eq:dLdS_3}
\end{align}
where the $\mathcal{L}_\text{emb}$ derivative w.r.t. $C_{\mu\tp}$ was already discussed in Appendix \ref{sec:orbital_response}. Summing up Eqs.~\ref{eq:dLdS_1}, \ref{eq:dLdS_2} and \ref{eq:dLdS_3} gives the full $\pdpd{\mathcal{L}_\text{emb}}{S_{\mu\nu}} S_{\mu\nu}^\RR$, and adding $\pdpd{\mathcal{L}_\text{emb}}{\boldsymbol{\mathcal{I}}}\boldsymbol{\mathcal{I}}^\RR$ (Eq.~\ref{eq:dLdI}) completes the $\pdpd{\mathcal{L}_\text{emb}}{\boldsymbol{A}} \boldsymbol{A}^\RR$ part of the DMET gradient in Eq.~\ref{eq:Lderiv}. The last piece, related to the mean-field response, is given in Eq.~\ref{eq:Lderiv_3}.

\bibliography{ref}

\end{document}